\documentstyle[epsfig,cite]{mn}

\begin{document}

\title[The optical long `period' of LMC~X-3]
    {The optical long `period' of LMC~X-3}
\author[Brocksopp et al.]
    {C.~Brocksopp$^{1,2}$\thanks{email: cb@astro.livjm.ac.uk}, P.J.~Groot$^{3,4}$, J.~Wilms$^5$\\
$^1$Astrophysical Research Institute, Liverpool John Moores University, 12 Quays House, Egerton Wharf, Birkenhead L41 1LD\\
$^2$Dept. of Physics \& Astronomy, The Open University, Walton Hall, Milton Keynes MK7 6AA\\
$^3$Astronomical Institute ``Anton Pannekoek'' and Center for High-Energy Astrophysics, University of Amsterdam,\\
\hspace{0.5cm}Kruislaan 403, 1098 SJ Amsterdam, The Netherlands\\
$^4$Harvard Smithsonian Center for Astrophysics, 60 Garden Street, Cambridge, MA 02138, USA\\
$^5$Institut f\"ur Astronomie und Astrophysik, Abt. Astronomie, Waldh\"auser Stra\ss e 64, D-72076 T\"ubingen, Germany\\}
\date{Accepted ??. Received ??}
\pagerange{\pageref{firstpage}--\pageref{lastpage}}
\pubyear{??}
\maketitle

\begin{abstract}
We have studied the long term variability of LMC X-3 in optical lightcurves spanning six years, in order to search for optical signatures which could confirm/refute the suggestion that the `modulation' is the result of accretion rate variability rather than accretion disc precession. We find that there is no stable period in the optical lightcurves, that the optical and X-ray lightcurves are correlated, with an {\em X-ray} lag and that there is no modulation in the optical colours. We argue that these effects agree with the variable mass accretion rate model proposed by Wilms et al. (2001).

\end{abstract}


\section{Introduction}
LMC X-3 was discovered by the {\it Uhuru} satellite (Leong et al. 1971), becoming headline news a few years later, when optical spectroscopy revealed an orbital period of 1.7 days and suggested that the compact object in the system was a black hole candidate (Cowley et al. 1983) -- at this time Cyg X-1 was the only other known BHC (see e.g. Tanaka \& Lewin 1995). Variability on a longer timescale ($\sim 100$ or 200 days) was discovered at both optical and X-ray wavelengths and precession of an accretion disc proposed as a likely mechanism (Cowley et al. 1991).

The spectral type of the companion star has been classified as B3 V (Cowley et al. 1983) and it is only recently that this has been questioned. Soria et al. (2001) used XMM Optical Monitor data to suggest that a B5 IV type is more likely and thus Roche lobe overflow can dominate the accretion onto the black hole. Given that the X-ray emission is typically soft and at high luminosity, Roche lobe overflow rather than wind accretion would indeed appear to be required.

With a high mass companion star, a black hole compact object and orbital and long term modulations on comparable timescales it was perhaps surprising that LMC X-3 and Cyg X-1 displayed such different X-ray properties. The `flickering' behaviour of the X-ray emission of Cyg X-1 and its associated hard power law energy spectrum had been thought to be potential black hole diagnostics. LMC X-3, on the other hand, had the much softer spectrum of an accretion disc and did not show the flickering behaviour. It was eventually pointed out by White \& Marshall (1984) that LMC X-3 was comparable with Cyg X-1 {\em when in the high/soft state}.

The more recent spectral state classification of black hole X-ray binaries (Van der Klis 1994, 1995) confirms that the `normal' state of LMC X-3 is the soft state. However, it was surprising that LMC X-3 was never observed in the hard state -- the state in which Cyg X-1 is almost always found (see e.g. Brocksopp et al. 1999). This was resolved recently by Wilms et al. (2001) with the discovery of a correlation between the X-ray hardness and the intensity -- indeed the minima of the `long modulation' were actually short ($\sim 1$ month) hard state periods. This hinted at variability of the mass accretion rate producing the observed long modulation, rather than precession of the accretion disc. Similarly the variability of the long period argues against accretion disc precession (Wilms et al. 2001; Paul, Kitamoto \& Makino 2000). The results of Wilms et al. (2001) have also been confirmed observationally by Boyd et al. (2000), Wu et al. (2000) and theoretically by Ogilvie \& Dubus (2000). 

\begin{figure*}
\begin{center}
\leavevmode  
\psfig{file=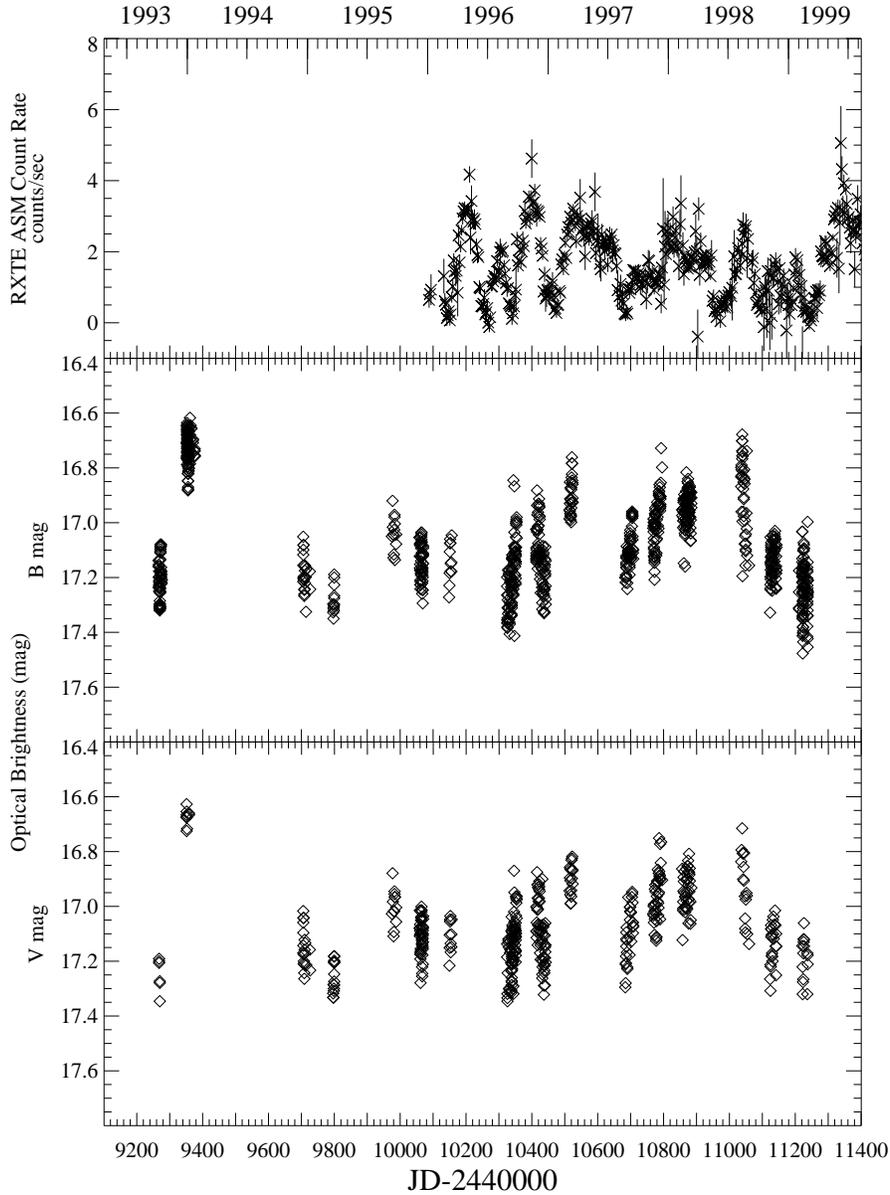,height=17cm,angle=0}  
\caption{$V$ and $B$ band photometry of LMC X-3 spanning 1993--1999. All available simultaneous X-ray data from the {\sl RXTE}/ASM are also plotted (daily averaged points).}
\label{lmcx3-lightcurve}
\end{center}
\end{figure*}

In this paper we study the optical behaviour of LMC X-3 in the $V$ and $B$ bands over a six year period in order to investigate any optical signatures which could indicate the nature of the long term variability of the lightcurves. Section~2 gives details of our observations and Section 3 looks at the orbital and long term variability, the relationship between the X-ray and the optical emission and the optical colours. Finally, we discuss our results in Section 4 and present our final conclusions in Section 5.

\section{Observations}

Data were obtained at the 0.91m Dutch Telescope of the European Southern Observatory in Chile; it was equipped with a 512$\times$512 TEK CCD and standard Johnson $V$ and $B$ filters. The observations were made between 1993 and 1999 in a total of sixteen different observing runs as part of the University of Amsterdam's monitoring program for X-ray binaries. Each observing run lasted 1--4 weeks, typically 2--12 observations of LMC X-3 would be made in a night and significantly more of the points were made in the $V$ band. Typical exposure times were five ($V$ band) and seven ($B$ band) minutes. Varying numbers of bias frames were taken, typically about ten at the beginning and ends of the night, and a total of 5--10 sky and dome flats each night.

All data were reduced as a single dataset using {\sc iraf}, following completion of the monitoring programme. Standard bias subtraction and flat field division were applied. {\sc iraf} was again used in order to obtain the photometry. Fifteen stars were selected on each frame, including LMC X-3. The mean FWHM was determined for each frame and this was then used to determine the radius of the aperture and the sky annulus -- the aperture was set to 1.5 times the FWHM, the inner radius of the surrounding annulus of sky was 4 times the FWHM and the width of the sky annulus was 5 times the FWHM.

Instrumental magnitudes for the fifteen stars were obtained and studied for stability over the six year monitoring campaign. In agreement with Van der Klis et al. (1985) we found that only two stars in the field (numbers 2 and 13 -- see Van der Klis et al. 1985 for details of these two stars and a finding chart) appeared to remain stable and thus be suitable reference stars for differential photometry. The two reference stars were calibrated with the {\sc iraf} tasks within {\sc photcal}, using photometric standard stars in the fields of PG 0231+051 and Mark A (Landolt 1992).Thus LMC X-3 could be calibrated against the results of the reference stars. As there was relatively little data for the standards, given the length of the monitoring campaign, the calibration of LMC X-3 depended heavily on the stability of the two reference stars used for the differential photometry over the time of the observations.
 
The calibrated dataset is plotted along with all available simultaneous {\sl RXTE}/ASM data (daily averaged points) in Fig.~\ref{lmcx3-lightcurve}.

\section{Results}

\subsection{The orbital lightcurves}
\begin{figure}
\begin{center}
\leavevmode  
\psfig{file=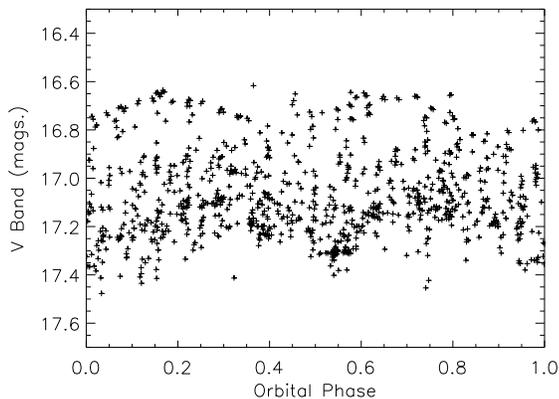,height=8cm,angle=90}  
\caption{The complete $V$ band dataset folded on the 1.7 day orbital period -- with such a large dataset the orbital modulation is not apparent due to the long term variability of the source.}
\label{fold}
\end{center}
\end{figure}

The complete dataset was initially folded on the 1.7 day orbital period, which has been well established in radial velocity measurements and also in the optical photometry of various authors (e.g. Van Paradijs et al. 1987). It was therefore quite surprising to find that the resultant plot was little more than scatter (Fig.~\ref{fold}). Clearly the large variability of the lightcurve makes it very hard to study the orbital modulation in a dataset spanning such a long period of time.

\begin{figure*}
\begin{center}
\leavevmode  
\psfig{file=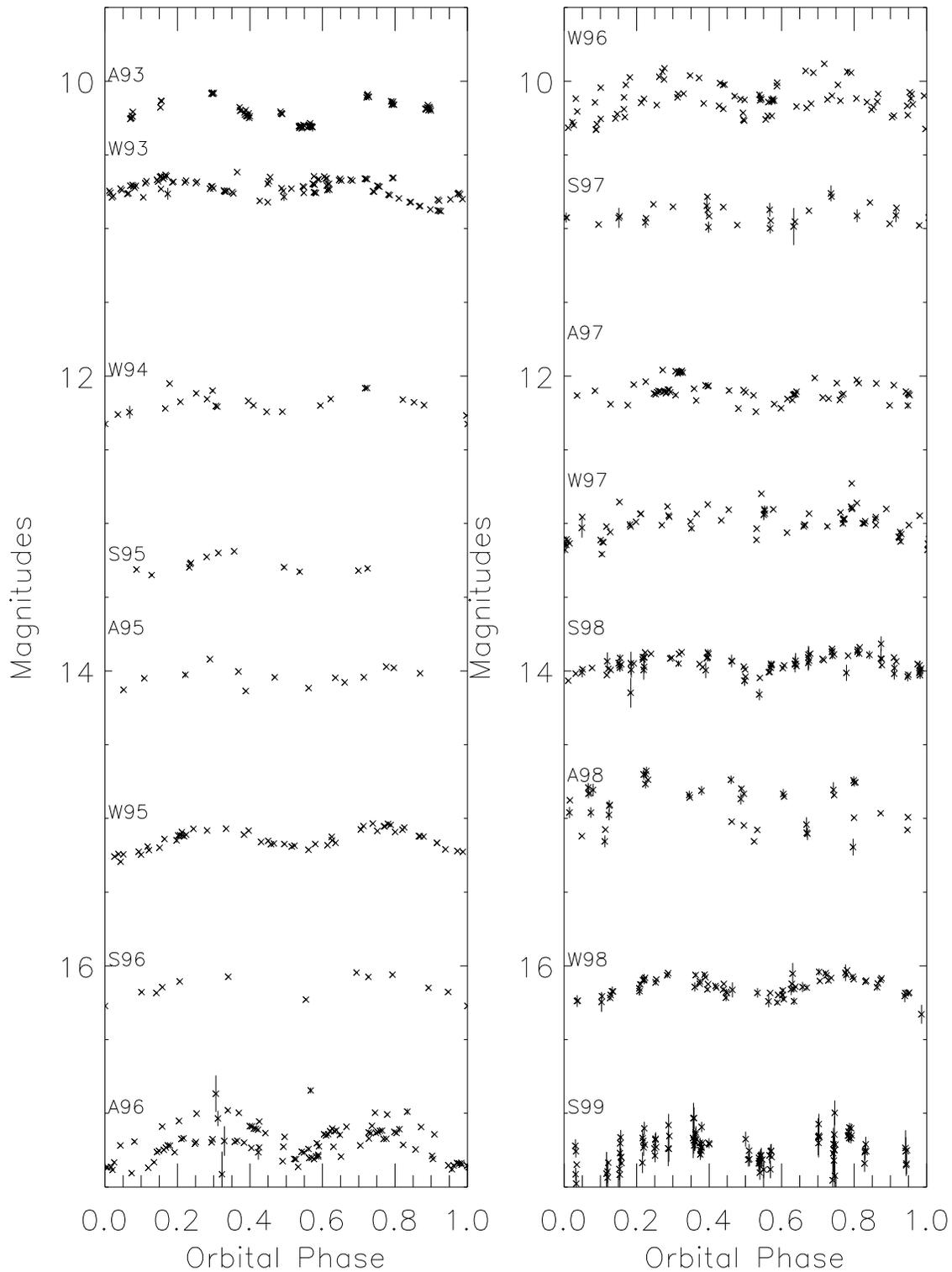,height=20cm,angle=0}  
\vspace*{1cm}
\caption{Mean orbital lightcurves in the $V$ band for the sixteen different observing sessions, each lasting 1--4 weeks. Successive observing sessions have been shifted by 1 magnitude for clarity and are labelled by a letter indicating spring, autumn or winter and a number indicating the year (e.g. S96 is spring 1996). Large amounts of variability can be seen from one session to the next and frequently within the few-week observing session (e.g. 1996).}
\label{orbital}
\end{center}
\end{figure*}

While it was perhaps tempting to follow the line of reseach of previous authors -- to split the data into sub-datasets, subtract the mean magnitude and then recombine into a smooth orbital curve -- as the ultimate aim of the project was to investigate the long modulation, this loss of information did not seem a productive way forward. It should also be noted that in the case of the precessing disc system LMC X-4, Heemskerk et al. (1989) find that on removal of the orbital period an orbital residual is left -- this is caused by the different heating effects caused at different angles of the precessing accretion disc. Reversing this effect shows that removing the long modulation will not result in a pure ellipsoidal curve and so we do not attempt it here.

The effect that the long modulation has on the orbital lightcurve can be seen in Fig.~\ref{orbital}. The dataset has been divided according to observing session, each session typically spans 1--4 weeks. The various sub-datasets have been folded on the 1.7 day orbital period and plotted, shifting successive plots by 1 magnitude. Each plot is labelled by a letter indicating spring, autumn or winter and a number indicating the year. In a number of the plots it can be seen that the two maxima/minima are not equal in magnitude, causing previous authors to assume that they are observing e.g. an accretion stream. However it is clear from Fig.~\ref{orbital} that the rise and fall of the lightcurve according to the `long period' is sufficiently steep to be responsible. While occasional sub-datasets show a smooth orbital lightcurve, it is more common for them to be scattered or to show a number of superimposed double-peaked curves. The latter half of 1996 shows clearly that within just 2--4 weeks the data from the different orbits becomes significantly shifted and thus incomparable without some method of `removing' the long modulation. Fitting the mean orbital curves with splines and comparing them shows that apparent phase shifts of up to 0.15 take place between the different observing sessions.

\subsection{Period Searching}
The study of the long modulation is therefore of much importance if the orbital lightcurves are ever to be understood properly; however, this is somewhat difficult on account of the lack of periodicity to the variability. Power spectra were computed for the photometry using the discrete Fourier transform routine in the {\sc starlink} package {\sc period}. The data were then split into ten orbital phase bins and the power spectra computed. In all cases there were no peaks in the power spectra that did not also appear in the window function of the Fourier transform. The Lomb-Scargle algorithm (also within {\sc period}, e.g. Scargle 1982) was used for comparison and similarly yielded no convincing modulation.

\begin{figure}
\begin{center}
\leavevmode  
\psfig{file=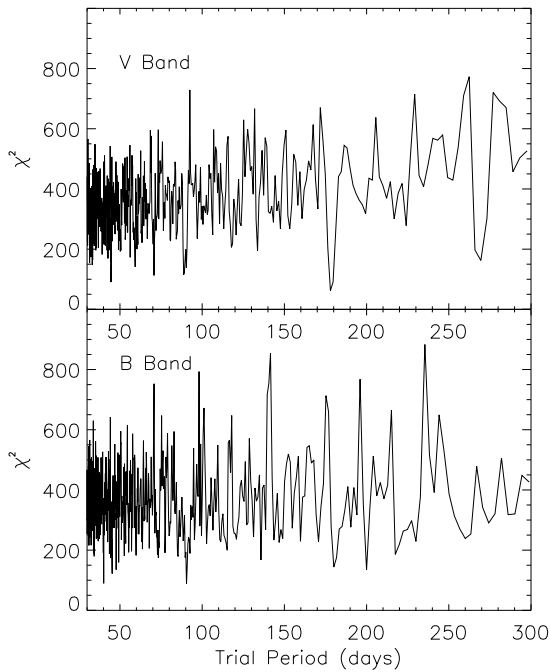,height=10cm,angle=0}  
\caption{The results of period-searching the complete $V$ and $B$ band datasets using the epoch-folding technique. Clearly there is no convincing period revealed by this method.}
\label{epoch}
\end{center}
\end{figure}

One reason for the lack of results in the power spectra could be the very intermittent sampling of the optical lightcurves (see Fig.~\ref{lmcx3-lightcurve}). The alternative method of `epoch-folding' was therefore used -- a technique which has been developed specifically for poorly sampled datasets. In this method the lightcurve is folded into $j$ phase bins on a range of different periods and the mean curves fit with straight lines (i.e. along the phase axis) -- if a period is present then there should be a clear peak in a plot of $\chi^2$ vs. period (for a time series of $N$ observations, $x_i$ ($i=1,2,...N$), folded into $M$ phase bins, $\chi^2=\sum_{j=1}^{M} (\bar{x}_j-\bar{x})^2/\sigma_j^2$ -- see e.g. Leahy et al. 1983). The data were treated as a whole and also as three sub-datasets, approximately 700 days in length although the sampling for each varied, and periods in the ranges 30--300, 30--150 and 150--300 days were tested --  unfortunately the results of this were equally inconclusive (see Fig.~\ref{epoch}).

The 198 day period found by Cowley et al. (1991) has already been found to vary slightly -- indeed Cowley et al. (1994) later suggested that it was actually half this value. More recently Wilms et al. (2001) reached similar conclusions from plotting a dynamical periodogram with the {\sl RXTE} data.

\begin{figure*}
\begin{center}
\hspace*{-3.5cm}\psfig{file=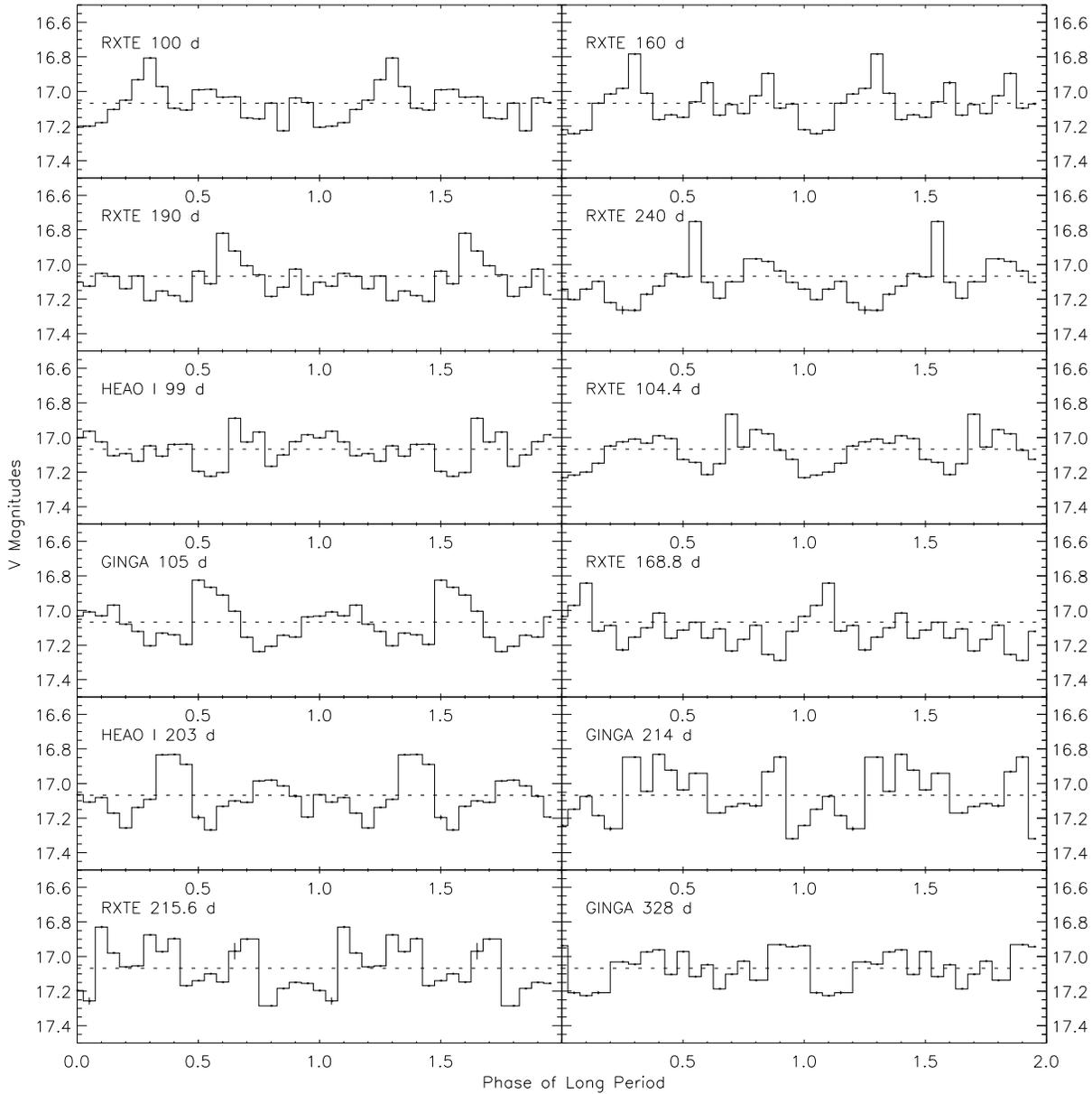,height=17cm,angle=0}  
\vspace*{.5cm}
\caption{The $V$ band photometric data folded on various `long periods' which have been reported in the literature. The top four plots are based on values quoted by Wilms et al. (2001); the bottom eight plots are based on values quoted by Paul, Kitamoto \& Makino (2000).}
\vspace*{0.5cm}
\label{Vfold_long}
\end{center}
\end{figure*}

Following the `periods' reported by Wilms et al. (2001) and Paul, Kitamoto \& Makino (2000) the optical lightcurves were folded on a selection of values and the resultant $V$ band plots shown in Fig.~\ref{Vfold_long}. In the top two plots of each column the optical data have been folded on values suggested by Wilms et al. (2001); the bottom part of the figure shows the optical data folded on periods reported by Paul, Kitamoto \& Makino (2000). Dotted lines across the plots show the mean magnitude of the complete dataset. The same value of JD$_0$ has been used for all plots (JD 2445626) -- this is somewhat arbitrary but is the closest time of minimum light (assuming a long period of $\sim100$ days) to the orbital ephemeris of Cowley et al. (1983).

After such unsuccessful period searching it is rewarding to see at least some evidence of the reported X-ray periods in the optical data with amplitudes significantly greater than $3\sigma$. Unfortunately the sampling seems to be complicating the results again as the $B$ and $V$ plots are not always consistent with each other. Note that there has been no provision made for the presence of the orbital period in the data (but see Section 3.4) and there is insufficient coverage for plots to be made for individual orbital phase bins. This line of research could be vastly improved by obtaining regularly sampled long term lightcurves in two or more optical filters, perhaps averaging into bins of length 1.7 days -- it would be an ideal project for a southern hemisphere robotic telescope.

\subsection{Comparison with the X-ray lightcurve}
As it is suspected that the optical and X-ray lightcurves are correlated it is perhaps surprising that the possible periodicities found in the X-ray lightcurves are not particularly apparent in the optical photometry. Similarly, it is also surprising that X-ray/optical flux--flux plots show only a slight correlation (Spearman rank correlation coefficient $\sim0.5$). While the variability of the long period and the presence of the orbital period in the optical data seem to be the main causes, it seems as though the sampling of the optical data may still be contributing to this and so cross-correlation of the X-ray and optical lightcurves was attempted in order to confirm this correlation.

\begin{figure}
\begin{center}
\leavevmode
\psfig{file=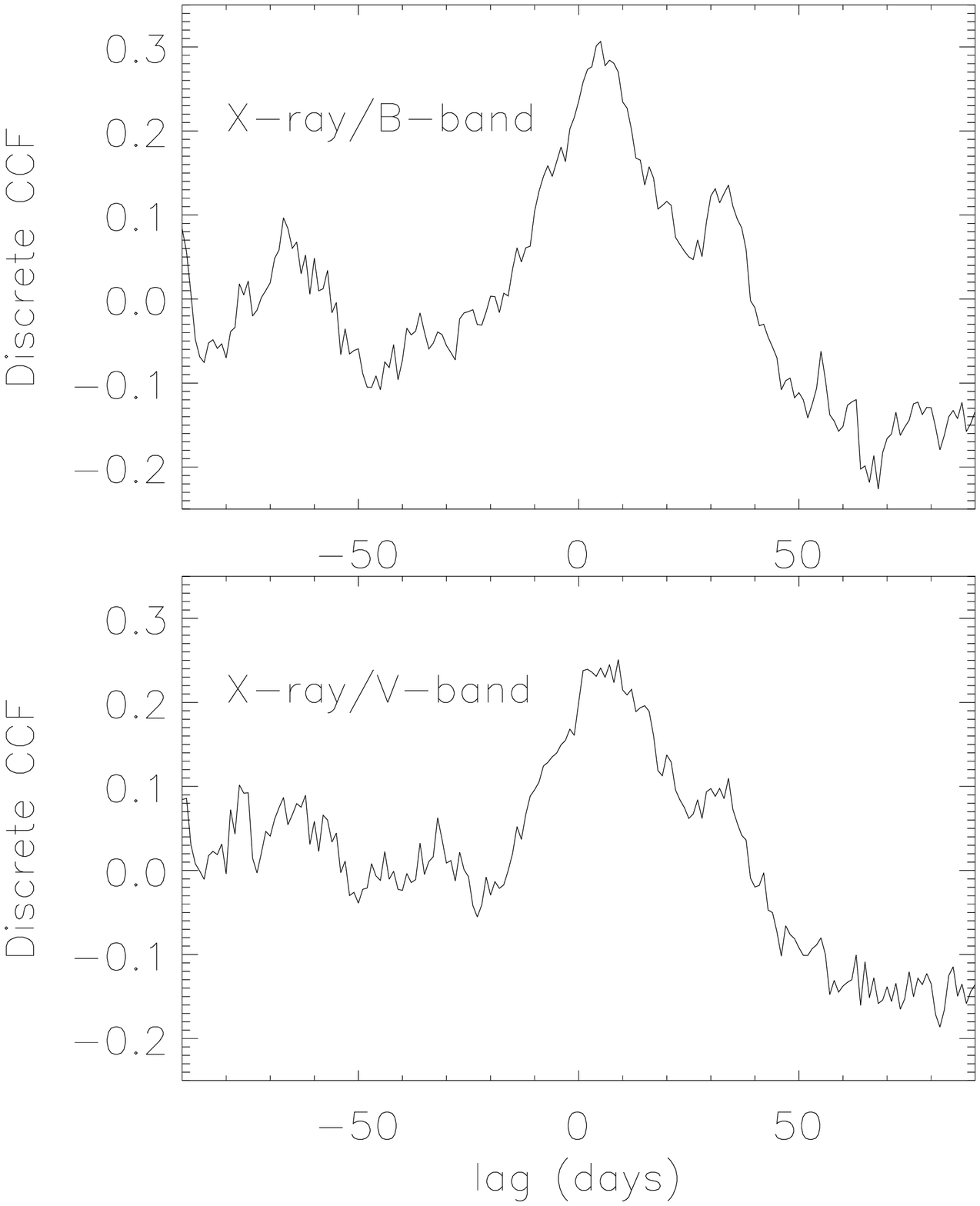,height=10cm,angle=0}  
\caption{The X-ray lightcurve cross-correlated with the $B$ (top) and $V$ band (bottom) lightcurves. A positive correlation and 5--10 day X-ray lag can be seen.}
\label{dccf}
\end{center}
\end{figure}

The discrete cross-correlation function (Edelson \& Krolik 1988) was obtained for each X-ray/optical pair and the resultant plots shown in Fig.~\ref{dccf} -- the optical magnitudes were converted into fluxes (mJy) (using the Johnson conversions) in order to do this. A clear peak can be seen in the centre of the plots, suggesting that the X-ray and optical lightcurves are indeed correlated. Furthermore, the X-ray lightcurve appears to lag the optical by $\sim$ 5--10 days -- this is approximately half of the lag found by Ebisawa et al. (1993), possibly suggesting variability in disc size or viscosity if the lag is related to the disc crossing time for the mass flow (see below). Unfortunately the time resolution of the data is insufficient to determine any lag between the $V$ and $B$ band lightcurves, although a $B$ band lag of $\sim 1$ day is hinted at. This X-ray lag would suggest that reprocessing of X-rays into optical emission or irradiation of the B star is {\em not} dominating the optical emission in LMC X-3; the correlation between the X-ray and optical would suggest that the ellipsoidal modulation of the star (that we would expect to see in the optical lightcurve but not in the X-rays) is not dominating the optical emission either. This is also clear to see from Figs.~\ref{fold} and \ref{orbital} where the amplitude due to the long variability is $\sim$ 2--3 times that of the orbital.

\begin{figure}
\begin{center}
\leavevmode  
\psfig{file=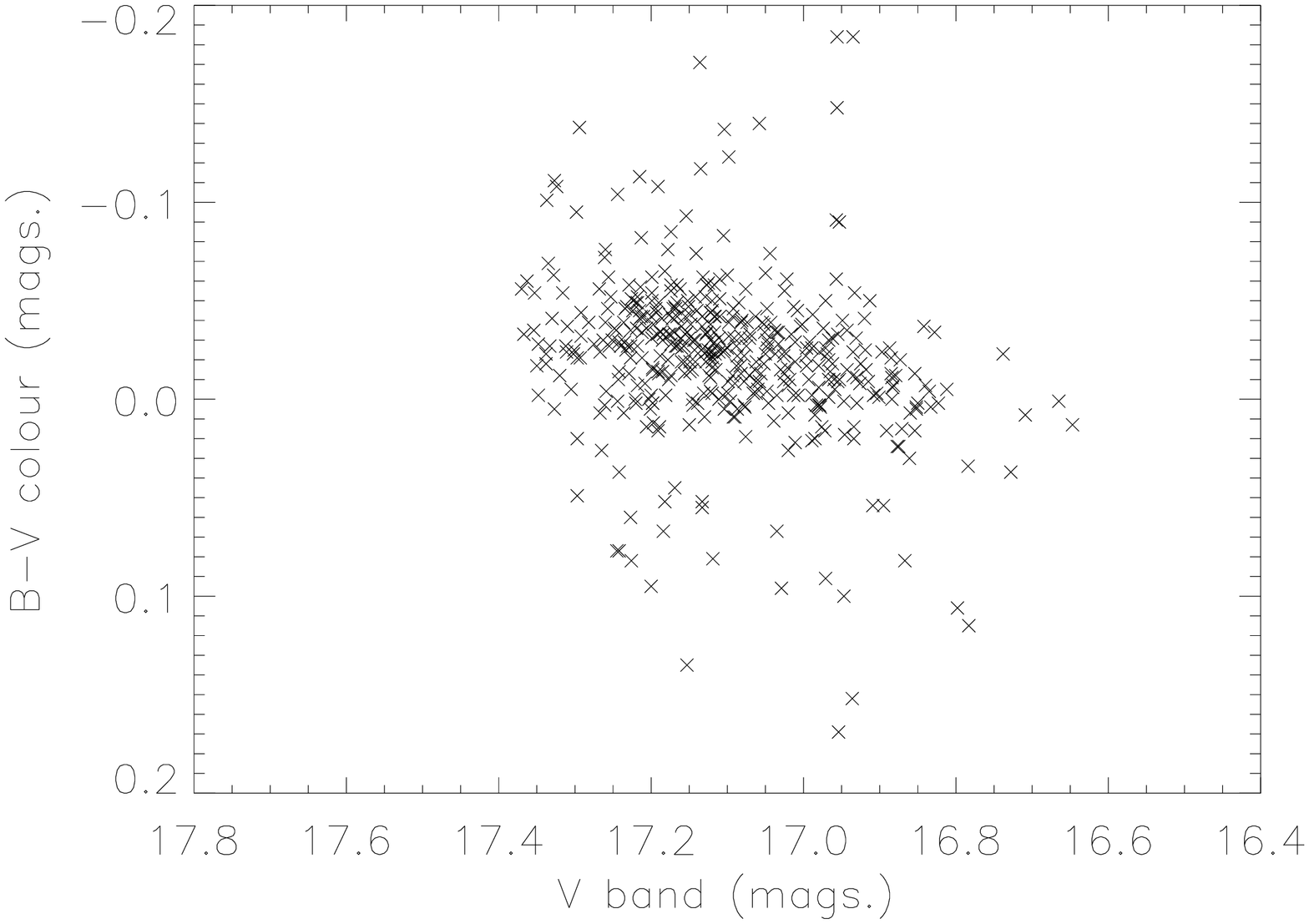,height=8cm,angle=90}  
\caption{The $B-V$ colours plotted against the $V$ band intensity. It appears that the correlation observed by Cowley et al. (1991) is less convincing in our data (but see text).}
\label{colour-corr}
\end{center}
\end{figure}

\subsection{The optical colours}


Despite a limited amount of $B$ band data, $B-V$ colours were determined where possible. Following Cowley et al. (1991) the $B-V$ data was plotted against the $V$ band -- the best fit straight line through the data points has a gradient of $-0.10\pm0.02$. Cowley et al. (1991) found correlated behaviour, in that the system became redder when the intensity increased, suggesting that the intensity increase was due to the brightening of a cool component, probably an accretion disc -- this correlation does not appear to be as convincing in our data, which obviously covers a much longer timescale (Fig.~\ref{colour-corr}). This is a somewhat strange result and it is unlikely that the intensity increases are wavelength independent -- however, we note that the scatter is most likely to be due to the presence of the orbital period, which Cowley et al. (1991) removed from their data; this is investigated further below.

To investigate the colour data further, the $B-V$ points were folded on the various long periods reported in the literature as for Fig.~\ref{Vfold_long} -- this time there was no evidence for a long modulation (with amplitude $>3\sigma$) at any of the suggested period values. Again, the implictions of this result are discussed below.

\subsection{To subtract the orbital period?}
Our study of the long modulation may be severely hampered by the presence of the orbital period, particularly as there is no 1.7 day modulation in the X-rays (although it may have been detected recently, Boyd 2001 private comunication). We therefore attempted to fit a double-peaked sine wave to the folded data of Fig.~\ref{fold} and thus subtract a model orbital lightcurve from the dataset. We then re-analysed our observations.

Folding the new optical dataset on the various values for the long period that have been suggested in the literature produced a plot that was essentially identical to that in Fig.~\ref{Vfold_long}. Likewise the correlation with the X-rays, including the lag, was unchanged. The correlation between the $V$ band and $B-V$ colours was improved on subtraction of this sine wave but still did not yield as convincing a correlation as found by other authors.

It should be noted that our best fit sine wave was found to have an amplitude of just $\sim 20\%$ that of Cowley et al. (1991). We do not suggest that the amplitude is variable (although this may be possible if the state changes producing the long variability result in significant changes in the geometry of the disc) but merely that the data in Fig.~\ref{fold} is too scattered to produce a reliable mean orbital lightcurve. Thus there is almost certainly some orbital signature remaining in our data and without significantly better time resolution we are unable to subtract it completely. 

However, we emphasize that even with improved time resolution, subtraction of the orbital period is non-straightforward. Fig.~\ref{orbital} illustrates how the long modulation does not only incline the mean orbital lightcurve upwards or downwards but also distorts the shape of the curve -- the asymmetry of e.g. the well-sampled curve of winter 1995 can be seen only too clearly.

\section{Discussion}
While the variation of a long modulation is not accounted for in traditional models of disc precession (e.g. Petterson 1977, Larwood 1998), more recent models including tilted/twisted discs due to coronal winds (e.g. Schandl \& Meyer 1994) and/or radiatively warped discs (e.g. Maloney \& Begelman 1997) can allow for these variations. Therefore, the fact that changes in the long period are observed in LMC X-3 does not rule out the possibility that the `modulation' is caused by a precessing disc. However, in a precessing system there should be a varying amount of obscuration by the disc and a hardening of the spectrum should be seen at low intensities (Paul, Kitamoto \& Makino 2000). They found that there is actually a weak positive correlation between the intensity and the 3$-$5 keV/1.5$-$3 keV hardness ratio which is therefore inconsistent with a precession model. Furthermore, the absorbing column density during the transition from soft to hard state does not increase (Wu et al. 2001). This indicates that the soft blackbody component seen in the X-ray spectra does not become obscured by warped parts of the disc. Paul, Kitamoto \& Makino (2000) concluded that given the multiple peaks present in the power spectrum (i.e. a fairly stable peak at $\sim 100$ days and a highly variable one at $\ge$ 130 days) it is possible that a combination of precession and mass accretion rate variability may be present.

The correlation between the X-ray and optical lightcurves again does not necessarily distinguish between precession and varying mass accretion rate; however, as we know that the X-ray source {\em changes spectral state} (Wilms et al. 2001) a varying mass accretion rate seems likely for both X-ray and optical emission, given the correlation. Furthermore, the 5--10 day X-ray lag does indeed rule out any suggestion of the optical emission being the result of reprocessing of X-rays. Were reprocessing responsible for the optical emission then we would expect to see optical delays of a few seconds instead (although obviously not in this dataset given the insufficient time resolution). The 5--10 day lag is most likely to represent the disc crossing time for increased mass flow passing through the disc, essentially the viscous timescale. One would normally expect the cooler, redder emission to come from the outer parts of the disc, with the hotter ultra-violet and X-ray emission found at smaller radii and so any new material entering the accretion flow would emit first in the $V$ band and (apparently) $\sim$ 5--10 days later reach the inner, X-ray emitting regions.

The correlation between the $V$ band and the $B-V$ colours found by previous authors and hinted at by our observations (although without sufficient subtraction of the orbital period we cannot confirm this) is consistent with this suggestion. If the disc is imagined to consist of a series of concentric rings emitting at X-ray, ultra-violet, $U$, $B$, $V$ and infrared wavelengths with progressively increasing disc radius, then a 5--10 day optical to X-ray delay would probably be consistent with a $\sim1$ day $V$ to $B$ band delay (cross-correlating the $B$ and $V$ lightcurves hints at this, although again we cannot be certain without removal of the orbital period).

Furthermore if the modulation were caused by a radiatively warped disc, as thought to be the case for e.g. Her X-1, then we might expect to see the same modulation in the optical colours. The presence of a modulation in the colours due to radiative warping would not be dependent on a short annulus crossing time (i.e. $\sim 1$ day) -- instead, additional radiation from warped parts of the accretion disc would contribute to the optical colours and `periodic' variability of significant amplitude (i.e. $\ge 3\sigma$) would be observed.

Alternatively, the presence of a flat, precessing disc could explain the lack of periodic colour variability -- the relative amounts of $V$ and $B$ emission would not change as different projected areas of the disc are observed. However, precession models require a rigid, tilted disc to tidally precess around the compact object and it is extremely unlikely that this would be possible; neither do the models explain what may tilt the disc in the first place. It is probable that disc precession is purely an idealised simplification of a warped, precessing disc and so it is unlikely that a flat precessing disc is present in LMC X-3.

In summary, the presence of a variable `long period', the hard state behaviour during the dips of the modulation (as found by Wilms et al. 2001), the delayed correlation between the X-ray and the optical lightcurves and the lack of variability of the $B-V$ colours are all suggestive that the long modulation of LMC X-3 is indeed due to a variable mass accretion rate. 

Such a modulation caused by a variable mass accretion rate could indeed take place, if a model of an accretion disc wind instability limit cycle is considered. Shields et al. (1986 and references within) showed that hard ($\ga$ 10 keV) X-rays emitted in the inner part of the accretion disc of X-ray binaries (and also quasars, AGN, CVs) will subject the outer parts of the disc to Compton heating, sufficient to drive a corona and possibly a wind from the disc. It was thought that such a wind could occur at a disc radius one tenth that at which the escape velocity equals the isothermal sound speed of the Compton-heated coronal gas. The model was extended to include attenuation and scattering of X-rays from the central source and the effects of inverse Compton cooling due to radiation emitted locally by the disc, thus enabling observational consequences of the phenomenon to be predicted and tested.

If the mass loss rate of the wind is small compared with the mass accretion rate then the wind-driving disc can remain in a steady state. However, if the disc luminosity and hence the mass loss rate of the wind exceed a critical value, then the mass of the disc and (following some delay) the mass accretion rate will decrease, reducing the luminosity of the X-ray source to a point at which it is insufficient to drive the Compton-heated wind. Thus a limit cycle is maintained. As the mass flow deficit passes radially through the disc on a viscous timescale it can be observed first in the optically emitting region of the disc and later in the inner X-ray emitting parts, producing the X-ray lag observed in LMC X-3. With the limit cycle requiring $\ga$ 90\% of the mass flow to be expelled from the system, the large amplitude oscillations of LMC X-3 can also be explained. 

An alternative model for the quasi-periodic X-ray state changes has been presented by Wu et al. (2001) and invokes a variable Roche lobe filling factor -- when the Roche lobe is filled then Roche lobe overflow can take place and the source is in a high luminosity soft state. When the Roche lobe becomes under-filled the mass accretion takes place via a focussed wind, causing a reduction in luminosity and subsequent hardening. While this may explain the X-ray observations and could still produce the X-ray lag presented in this paper, it is not clear how the Roche lobe alternates between being filled and under-filled. Our observations are currently unable to distinguish between the two suggested models -- further optical/ultraviolet spectroscopic analysis is in progress in order to determine characteristics of the stellar wind (Brocksopp et al. in prep.).

A way in which these results could be tested would be to obtain simultaneous high time resolution ($\sim$ hours) X-ray, ultra-violet, optical and infrared data over the course of a cycle of variability -- this would enable much more accurate timing analysis than possible with the dataset presented here. Further confirmation of the hard states in the dip of the modulation could be obtained from the discovery of a radio jet, as seen in other X-ray binaries during the hard state (see e.g. Fender 2001). The long modulation of Cyg X-1 (see Brocksopp et al. 1999) should also be re-examined in the context of the Compton-heated coronal wind model as a number of authors have thought it an unlikely pure-precession candidate (e.g. Ogilvie \& Dubus, 2000).

\section{Conclusions}
The long term behaviour of LMC X-3 has been studied in the optical photometry, in conjunction with the recent discovery of low/hard X-ray spectral states. No stable periods have been found and this is consistent with analysis of the X-ray lightcurves. The correlation between the X-ray and optical data and also the X-ray delay would suggest that increased mass flow through the disc is observed as it passes through the cold outer regions into the hot inner regions -- thus it is likely that the long term variability is produced by changes in mass accretion rate, rather than by the precession of the accretion disc. This could be confirmed by acquiring optical data with better time resolution and the inclusion of ultraviolet observations.

\section*{acknowledgements}
We are extremely grateful to the University of Amsterdam whose X-ray binary monitoring programme provided the data for this project. We would like to thank Paul Vreesrijk for all his help with the data reduction and Phil Charles, Andy Norton, Martin Heemskerk and Roberto Soria for useful comments. CB acknowledges a PPARC studentship whilst at the Open University and notes that much of this work was completed on the Sussex STARLINK node.

\end{document}